\def\nobs{\eta_{\rm 0}}
\def\ne{\eta_{\rm ls}}
\def\nend{\eta_{\rm end}}
\def\nq{\eta_{\rm eq}}
\def\al{\langle a_l^2\rangle}
\def\zeq{Z_{\rm eq}} 
\def\zls{Z_{\rm ls}} 
\def\zend{Z_{\rm end}}
\def\ed{{\rho_{\rm dS}\over\rho_{P}}}
\def\A{{\cal A}}
\def\B{{\cal B}}
\def\C{{\cal C}}
\def\S{{\sf S}}
\def\T{{\sf T}}

\documentstyle[eqsecnum,preprint,aps,prd]{revtex} 
\tighten
\begin{document}
\preprint{WISC-MILW-94-TH-18} 
\draft

\title{CBR Anisotropy from Primordial Gravitational Waves in
Two-component Inflationary Cosmology}

\author{Scott Koranda}
\address{
Department of Physics, University of Wisconsin -- Milwaukee\\
P.O. Box 413, Milwaukee, Wisconsin 53201, U.S.A.\\
email: skoranda@dirac.phys.uwm.edu}
\author{Bruce Allen}
\address{
Department of Physics, University of Wisconsin -- Milwaukee\\
P.O. Box 413, Milwaukee, Wisconsin 53201, U.S.A.\\
email: ballen@dirac.phys.uwm.edu}
\date{\today}
\maketitle
\begin{abstract}
We examine stochastic temperature fluctuations of the cosmic background
radiation (CBR) arising via the Sachs-Wolfe effect from gravitational
wave perturbations produced in the early universe. We consider
spatially flat, perturbed FRW models that begin with an inflationary
phase, followed by a mixed phase containing both radiation and dust.
The scale factor during the mixed phase takes the form
$a(\eta)=c_1\eta^2+c_2\eta+c_3$, where $c_i$ are constants.  During
the mixed phase the universe smoothly transforms from being radiation
to dust dominated. We find analytic expressions for the graviton mode
function during the mixed phase in terms of spheroidal wave functions.
This mode function is used to find an analytic expression for the
multipole moments $\langle a_l^2\rangle$ of the two-point angular
correlation function $C(\gamma)$ for the CBR anisotropy.  The analytic
expression for the multipole moments is written in terms of two
integrals, which are evaluated numerically. The results  are compared
to multipoles calculated for  models that are {\it completely} dust
dominated at last-scattering. We find that the multipoles $\langle
a_l^2\rangle$ 
of the CBR temperature perturbations
for $l>10$ are  significantly larger for a universe that
contains both radiation and dust at last-scattering.  We compare our
results with recent, similar numerical work and find good agreement.
The spheroidal wave functions may have applications to other problems
of cosmological interest.
\end{abstract}


\section{INTRODUCTION}
\label{intro}

This paper considers the effect of primordial gravitational waves on
the cosmic background radiation (CBR). We consider spatially flat,
perturbed Friedman-Robertson-Walker (FRW)
universes that begin with an early inflationary phase. As
the universe rapidly expands, perturbations of the spatial geometry
that are local in origin (eg.  thermal fluctuations) are quickly
redshifted both in amplitude and wavelength.  After sufficient
inflation these perturbations are no longer visible to an observer; the
only perturbations that remain visible within a Hubble sphere are
quantum-mechanical zero-point fluctuations.  Because these
perturbations extend  to arbitrarily high frequencies they can not be
redshifted away. Since the only significant perturbations remaining
after inflation are zero-point fluctuations, we assume that the initial
state of the universe was the vacuum state appropriate to de Sitter
space, containing only the quantum fluctuations and no additional
excitations.  As the universe continues to expand after inflation,
these quantum fluctuations are redshifted to longer wavelengths and
amplified; one may think of this in terms of particle (graviton)
production (as we do), non-adiabatic amplification, or super-radiant
scattering. In the present epoch the occupation numbers of the graviton
modes are large.  This is not surprising.  It has been shown that one 
may treat the collective effects of primordial gravitational waves as being due to
the presence of a stochastic background of classical gravitational
waves, and since gravitons are bosons, such an interpretation is only
possible if the occupation numbers are large. 

Sachs and Wolfe \cite{SachsWolfe} showed how gravitational wave
perturbations result in CBR temperature anisotropy. As photons from the
CBR propagate,  the paths they follow are perturbed by the metric
perturbation $h_{ij}$, which in this discussion is due entirely to
primordial gravitational waves. The energies of the photons are
perturbed, which results in  temperature fluctuations from point to
point on the celestial sphere. These temperature fluctuations are
usually characterized by the two-point angular correlation function
$C(\gamma)$ defined on the celestial sphere.  Here $\gamma$ is the
angle between two points on the sphere. Most often the two-point
angular correlation function is expanded in terms of Legendre
polynomials, and the expansion coefficients or multipole moments are
calculated. For a derivation of the angular correlation function for
spatially flat cosmologies see the recent paper by Allen and Koranda
\cite{AllenKoranda}. Henceforth we will assume that the reader
is familiar with this paper, which contains a detailed review of previous
work on this problem, a comprehensive discussion of the physical
motivation, and a detailed and self-contained ``first-principles''
calculation.

Early work on this subject \cite{Rubakov,Fabbri,AbbotWise,Starobinsky}
assumed that the universe was completely dust dominated at
last-scattering when the CBR decoupled, as did recent work
by White \cite{White1} which presented a concise derivation of the
formula for the multipole moments due to tensor perturbations. Also
recently, Grishchuk has adapted the terminology and techniques of
quantum optics to the analysis \cite{Grishchuk}. Grishchuk stresses the
importance of the phase correlations between the modes of the metric
perturbations.  A similar analysis by Allen and Koranda
\cite{AllenKoranda}   using standard ``quantum field theory in curved
space'' techniques found equivalent results.  We also showed that the now
standard formula given by Abbott and Wise \cite{AbbotWise} and
Starobinsky \cite{Starobinsky} for the $l$'th multipole moment is a
long wavelength approximation to an exact formula. Both the work by
Grishchuk, and by Allen and Koranda assumed that the universe was
completely dust dominated at last-scattering.

The first authors to consider the CBR anisotropy (within the framework of the
Sachs-Wolfe effect) for a universe that was {\it not} completely
dust dominated at last-scattering were Turner, White, and Lidsey
\cite{Turner2}.  They used a ``transfer function'' to express the
solutions of the wave equation for the gravitational wave amplitude in
terms of the standard long wavelength formula given by Abbott and
Wise.  They found that the standard formula, which assumes that the
universe is {\it completely} dust dominated at last-scattering,
consistently underestimates the contribution of gravitational waves to
the CBR anisotropy.  More recent work by Ng and Speliotopoulos
\cite{Ng} used numerical methods to integrate the wave equation and
found similar results.  Neither of these studies used analytic
expressions for the gravitational wave amplitude (or equivalently, the
graviton mode function).

Other work has been done which does not directly use the Sachs-Wolfe formula to
calculate the anisotropy due to gravitational waves. Crittenden et.
al.  \cite{Crittenden} numerically evolved  the photon distribution
function using first-order perturbation theory of the general
relativistic Boltzmann equation for radiative transfer, and included a
Thomson scattering source term. Dodelson, Knox and Kolb \cite{KnoxDodelson}
have done a similar numerical analysis. Both found that the standard
formula of Abbott and Wise is only accurate for small $l$ multipole
moments, and consistently underestimates the contribution of the
gravitational waves to the CBR anisotropy for higher $l$ moments.

In this paper we give the first correct analytic expression for the
graviton mode function in a cosmology that transforms smoothly from
being radiation  to dust dominated. (We correct a minor error in
earlier work by Sahni \cite{Sahni} and Nariai \cite{Nairi}  which
claimed to find an analytic expression for the mode function.) We use
this analytic expression and the Sachs-Wolfe formula to find an
analytic expression for the multipole moments $\langle a_l^2\rangle$ of
the angular correlation function $C(\gamma)$.  The analytic expression
for the multipole moments is written in terms of two integrals; we use
numerical methods to evaluate these integrals and report numerical
values for the multipole moments.  We compare our results with those
mentioned above, and find good agreement.

The paper is organized as follows.  In section
\ref{angularcorrelationfunction} we reproduce general expressions for the
angular correlation function derived in  \cite{AllenKoranda}, and
explain how one uses these formulae to calculate the multipole moments
for any spatially flat, inflationary cosmological model. In section
\ref{cosmologicalmodel} we introduce our cosmological model, which
``begins'' with an infinite de Sitter phase followed by a mixed phase
that contains both radiation and dust. Early in the mixed phase the
universe is radiation dominated; later it transforms smoothly from
being radiation dominated to dust dominated. In section
\ref{gravitonmodefunction} we solve the massless Klein-Gordon equation
(or wave equation) for the graviton mode function. During the mixed
phase the solutions to the wave equation are expressed in terms of
spheroidal wave functions. The multipole moments are calculated in
section \ref{multipolemoments} using the graviton mode function
determined in section \ref{gravitonmodefunction}.  An analytic
expression  for the moments is given in terms of two integrals, which
are then evaluated numerically.  In appendix \ref{appendixa}
we discuss the spheroidal wave functions. The differential equation of
spheroidal wave functions is introduced and its solutions examined. We
introduce a useful notation for the spheroidal wave functions, and give
a practical method for  evaluating them. Finally in appendix \ref{appendixb}
we describe the numerical techniques used to evaluate the 
spheroidal wave functions
and the two multipole moment integrals.

Throughout this paper we use units where the speed of light
$c=1$. We  retain Newton's gravitational constant
$G$ and Planck's constant $\hbar$ explicitly.


\section{The Angular Correlation Function}
\label{angularcorrelationfunction}

We consider only the anisotropy of the cosmic background
radiation (CBR) 
arising via the Sachs-Wolfe effect \cite{SachsWolfe} from
tensor perturbations (gravitational
waves). 
The anisotropy is characterized by the two-point angular
correlation function $C(\gamma)$, where $\gamma$ is
the angle between two points located on the celestial sphere.
The correlation function
may be expanded in terms of Legendre Polynomials as
\begin{equation}
C(\gamma)=\bigg\langle {\delta T\over T}(0){\delta T\over T}(\gamma)
\bigg\rangle=\sum_{l=0}^\infty
{(2l+1)\over 4\pi} \langle a_l^2\rangle P_l(\cos\gamma).
\label{correlationdefinition}
\end{equation}
The expansion coefficients $\langle a_l^2\rangle$ are 
referred to as the multipole moments. The multipole moments
are given in terms of an integral over graviton wavenumber
$k$,
\begin{equation}
\langle a_l^2\rangle\equiv 4\pi^2(l+2)(l+1)l(l-1)\int_0^\infty
{dk\over k}|I_l(k)|^2,
\label{aldefinition}
\end{equation}
where the function $I_l(k)$ is proportional to the Sachs-Wolfe
integral along null geodesics, and is given by
\begin{equation}
I_l(k)\equiv \int_0^{\nobs-\ne} \> d\lambda\> F(\lambda,k) {j_l(k(\nobs
-\ne-\lambda))\over k(\nobs-\ne-\lambda)^2}.
\label{Ildefinition}
\end{equation}
Here  $\ne$ is the time of last-scattering, and $\nobs$ is the
conformal time today. The function $j_l(z)$ is a spherical
Bessel function of the first kind \cite{Jackson}.
The function $F(\lambda,k)$ is proportional
to the first derivative of the graviton mode function $\phi(\eta,k)$,
and is defined by
\begin{equation}
F(\lambda,k)\equiv k^{1/2}\bigg[ {\partial\over\partial\eta}
\phi(\eta,k)\bigg]_{\eta=\ne+\lambda}.
\label{Fdefinition}
\end{equation}
The graviton mode function obeys the massless Klein-Gordon
or wave equation
\begin{equation}
\ddot\phi +2{{\dot a}(\eta)\over a(\eta)}\dot\phi
+ k^2\phi=0,
\label{kge}
\end{equation}
where $a(\eta)$ is the scale factor and
\begin{equation}
\cdot\equiv {\partial\over\partial\eta}.
\end{equation}
The mode function must satisfy the Wronskian normalization
condition
\begin{equation}
\big\{\phi(\eta,k)\dot{\phi^*}(\eta,k)
-\phi^*(\eta,k)\dot\phi(\eta,k)\big\} =-{2 i\hbar G\over \pi^2
a^2(\eta)}.  
\label{phinormalization} 
\end{equation}
The only physical input required is the choice of an initial
quantum state for the gravitational field, which amounts to a choice of
boundary conditions for the wave equation (\ref{kge}).

These formula for the angular correlation function are very general; a
detailed and complete derivation is given in \cite{AllenKoranda}.  To
calculate the correlation function (or equivalently the multipole
moments) for any particular cosmological model, one need only to solve
the wave equation (\ref{kge}) for the graviton mode function, and
substitute into the formulae
(\ref{correlationdefinition}-\ref{Fdefinition}).


\section{The Cosmological Model}
\label{cosmologicalmodel}

The spacetime considered here is a spatially flat, perturbed FRW
universe. The metric  is 
\begin{equation}
ds^2=a^2(\eta)[-d\eta^2+\big(\delta_{ij}+h_{ij}(\eta,x^k )\big) dx^i
dx^j],
\label{metric}
\end{equation}
where $\delta_{ij}$ is the flat metric of $R^3$, $\eta$ is the
conformal time, and $a(\eta)$ is the cosmological  length scale
or scale factor. The scale factor satisfies the Einstein equations
\begin{equation}
{{\dot a}^2(\eta)\over a^4(\eta)}={8\pi G\over 3}\rho(\eta)\>\>
\text{and}\>\>{{\ddot a}(\eta)\over a^3(\eta)}-
{{\dot a}^2(\eta)\over a^4(\eta)}=-{4\pi G\over 3}\Big(
\rho(\eta)+3P(\eta)\Big),
\label{Einstein}
\end{equation}
where $\rho(\eta)$ is the energy density and $P(\eta)$
the pressure of the cosmological fluid.
The metric perturbation $h_{ij}$ is assumed to be
small; in the limit as $h_{ij}$ vanishes the spacetime is an
unperturbed FRW universe. We have chosen a gauge so that the tensor
perturbation $h_{ij}$ has only spatial components. 

In the absence of the metric perturbation $h_{ij}$, the background
spacetime is a spatially flat FRW universe. Since the
spacetime is spatially flat,
the density parameter $\Omega_0$,
which is the ratio of the present-day energy density $\rho_0$
to the critical energy density required to produce
a spatially flat universe, is equal to unity:
\begin{equation}
\Omega_0={8\pi G\rho_0\over 3 H_0^2}=1.
\end{equation}
To  specify the  model, we need to give the
scale factor $a(\eta)$.  We  do this in such
a way that the model is completely defined by 
the minimal set of free parameters given
in Table \ref{table1}. 
{\it All} our
results, including the final expression (\ref{moments}),
can be expressed in terms of this minimal set of parameters. For
clarity,  we often define auxiliary quantities and express
results in terms of them; the auxiliary quantities can be expressed
in terms of the parameters in Table
\ref{table1}. In typical inflationary models, the free
parameters in Table \ref{table1} have values of order $H_0$ between 50 and 100
km $\rm{s}^{-1}$ $\rm{Mpc}^{-1}$, $100<\zls<1500$, $2\times 10^3<\zeq<2\times 10^4$, and
$10^{20} <\zend$. For a review of inflationary cosmology see
reference \cite{KolbTurner}.

\subsection{The Inflationary or de Sitter Phase}

Our cosmological model passes through two phases. 
The first  phase is a de Sitter or inflationary phase.
During the de Sitter phase the universe expands
exponentially (expressed in terms of
{\it comoving} time $t$, the scale factor behaves like 
$a(t)\sim e^{Ht}$, where $H$ is the Hubble constant).
In terms of conformal time, the scale factor is 
\begin{equation}
a(\eta)=a(\nend)\bigg(2-{\eta\over\nend}\bigg)^{-1}\>\>
\text{for}\>\>\eta\leq\nend,
\label{deSitterscalefactor}
\end{equation}
where $\nend$ is the conformal time at the end of
the de Sitter phase. In terms of the  parameters
listed in Table \ref{table1}, 
\begin{eqnarray}
a(\nend)&=&(1+Z_{\rm end})^{-1},\\
&&\nonumber\\
\nend&=& H_0^{-1}\Bigg[{(2+\zeq)\over (2+\zeq+\zend)
(1+\zend)}\Bigg]^{1/2}\approx H_0^{-1}{\sqrt{\zeq}\over  \zend},
\end{eqnarray} 
where we have set the scale factor today $a(\eta_0)=1$.
During the de Sitter phase the energy density $\rho_{\rm dS}$
is  constant and given by
\begin{equation}
\rho_{\rm dS}={3\over 8\pi G}{{\dot a}^2(\nend)\over
a^4(\nend)}=
{3 H_0^2 \over 8\pi G} {(1+Z_{\rm end})^3\over
(2+Z_{\rm eq})} (2+Z_{\rm eq}+Z_{\rm end})\approx
{3\over 8\pi G} H_0^2 
{\zend^4\over\zeq}={\zend^4\over \zeq}\rho_0.
\end{equation}
The pressure during the de Sitter phase
is negative and constant, and is given by
\begin{equation}
P_{\rm dS}=-\rho_{\rm dS}.
\end{equation}
The de Sitter phase is followed by a mixed phase.

\subsection{The Mixed Radiation and Dust Phase}

Immediately following the de Sitter phase is a mixed phase
containing both dust and radiation. The scale factor is
\begin{equation}
a(\eta)=a(\nend)\bigg[{1\over 4}\bigg({\xi\over 1+\xi}\bigg)
\bigg({\eta\over\nend}-1\bigg)^2+{\eta\over\nend}
\bigg]
\>\>\text{for}\>\>\eta\geq\nend.
\label{mixedscalefactor}
\end{equation}
The constant $\xi$  is defined in terms of the free
parameters by
\begin{equation}
\xi={1+\zeq\over 1+\zend}.
\label{xidefinition}
\end{equation}
The stress-energy tensor is that of
a perfect fluid with energy density 
\begin{equation}
\rho(\eta)=\rho_{\rm dust}(\eta)+\rho_{\rm rad}(\eta)
\>\>\text{for}\>\>\eta\geq\nend,
\label{mixedenergydensity}
\end{equation}
where the energy density for the dust $\rho_{\rm dust}(\eta)$,
and for the radiation $\rho_{\rm rad}(\eta)$, are 
\begin{eqnarray}
\rho_{\rm dust}(\eta)&=&{\rho_{\rm eq}\over 2}{a^3(\nq)\over a^3(\eta)},\\
\rho_{\rm rad}(\eta)&=&{\rho_{\rm eq}\over 2}
{a^4(\nq)\over a^4(\eta)}\label{radiationenergydensity}.
\end{eqnarray}
Here $\rho_{\rm eq}=\rho(\eta_{\rm eq})$ 
is the energy density at conformal time $\nq$, when the
dust and radiation energy densities are equal, and is given
by 
\begin{equation}
\rho_{\rm eq}={3 H_0^2\over 4\pi G}
{(1+Z_{\rm eq})^4\over (2+Z_{\rm eq})}
\approx 2\bigg({\zeq\over\zend}\bigg)^4\rho_{\rm dS}\approx 2\zeq^3 \rho_0.
\end{equation}
The pressure of the cosmological fluid $P(\eta)$ is 
\begin{equation}
P(\eta)=P_{\rm rad}(\eta)= 
P_{\rm eq}{a^4(\nq)\over a^{4}(\eta)} \>\>\text{for}
\>\>\eta\geq\nend,
\label{mixedpressure}
\end{equation}
where the pressure at dust/radiation equality $P_{\rm eq}$ is
\begin{equation}
P_{\rm eq}=P(\eta_{\rm eq})={\rho_{\rm eq}\over 6}.
\label{pressureeq}
\end{equation}
By inspection of (\ref{radiationenergydensity}),(\ref{mixedpressure}),
and (\ref{pressureeq}) one sees that
\begin{equation}
P(\eta)=P_{\rm rad}(\eta)={1\over 3}\rho_{\rm rad}(\eta),
\end{equation}
which is the relation between the pressure and energy density
one would expect since the dust has zero pressure and the pressure is
due entirely to the radiation present.
The scale factor at $\nq$ is
\begin{equation}
a(\nq)=(1+Z_{\rm eq})^{-1}.
\end{equation}
One can express $\nq$ 
in terms of the conformal time at the
end of the de Sitter phase $\nend$ and the constant $\xi$ as
\begin{equation}
\nq={\nend\over\xi}(2\sqrt{2(1+\xi)}-2-\xi)
\approx H_0^{-1}{2(\sqrt{2}-1)\over \sqrt{\zeq}}.
\end{equation}
For $\eta\ll\nq$ the energy density (\ref{mixedenergydensity})
varies as $\rho(\eta)\sim a^{-4}(\eta)$ and
the cosmological model is radiation dominated, and for $\eta\gg\nq$ 
the energy density varies as $\rho(\eta)\sim a^{-3}(\eta)$ and the model
is dust dominated. The model makes a smooth transition 
at $\eta=\nq$ from being radiation to dust dominated.


\section{The Graviton Mode Function}
\label{gravitonmodefunction}

To calculate the multipole moments for the cosmological
model described above, one must first solve the wave
equation (\ref{kge}) for the graviton
mode function. We first solve for the ``natural''
positive- and negative-frequency mode functions during both the de Sitter and
the mixed phases. We then make an appropriate choice of initial
mode function during the de Sitter
phase. This choice of initial mode function completely
determines the graviton mode function for all later
times. We express the mode function at later times
using Bogolubov coefficient notation.

\subsection{The de Sitter Phase}

The solution to the wave equation (\ref{kge}) for the
de Sitter phase can be expressed in terms of spherical
Hankel functions. The scale factor during the de Sitter
phase is given in (\ref{deSitterscalefactor}). By making the change of dependent
variable 
\begin{equation}
\chi(\eta,k)=(\eta-2\nend)^{-2}\phi(\eta,k),
\end{equation}
and the change of independent variable
\begin{equation}
z=k(\eta-2\nend),
\end{equation}
the wave equation can be expressed in the form
\begin{equation}
{d^2\chi\over d z^2}+{2\over z}{d\chi\over d z}
+\bigg(1-{2\over z^2}\bigg)\chi=0.
\end{equation}
This is Bessel's differential equation, and the solutions
are spherical Bessel or Hankel functions  
\cite{Jackson}.  Using the normalization
condition (\ref{phinormalization})
one obtains for the positive-frequency mode function during
the de Sitter phase
\begin{equation}
\phi^{(+)}_{\rm ds}(\eta,k)=-i\sqrt{{8\over 3\pi}\ed }\>e^{-ik\nend}
\>\nend^2 k^{1/2}
{a^2(\nend)\over a^2(\eta)} h^{(2)}_1\Big(k(\eta-2\nend)\Big),
\label{phidesitter}
\end{equation}
where $h_1^{(2)}(z)$ is a spherical Hankel function of
the second kind \cite{Jackson}, and 
$\rho_{P}=1/\hbar G^2\approx 5\times 10^{93} {\rm gm}
/{\rm cm}^3$ is the Planck energy density. 
The negative-frequency mode function during the de Sitter
phase $\phi_{\rm ds}^{(-)}(\eta,k)$ is just the complex
conjugate of the positive-frequency mode function 
(\ref{phidesitter}). The positive- and negative-frequency
mode functions form a complete solution to the wave
equation for the de Sitter phase.

\subsection{The Mixed Radiation and Matter Phase}
\label{mixed}

The  wave equation during the mixed radiation and dust phase 
$(\eta>\nend)$ can be
cast in the form of the spheroidal wave function differential
equation.  By making the change of dependent variable
\begin{equation}
\chi(\eta,k)\equiv a^{1/2}(\eta)\phi(\eta,k),
\label{chidefinition}
\end{equation}
the wave equation (\ref{kge}) becomes
\begin{equation}
{\ddot \chi}+{\dot a(\eta)\over a(\eta)}\dot\chi+
\bigg(k^2-{1\over 4}{{\dot a}^2(\eta)\over a^2(\eta)}
-{1\over 2}{\ddot a(\eta)\over a(\eta)}\bigg)\chi=0.
\label{wavetwo}
\end{equation}
One may define a new independent variable
\begin{equation}
x\equiv\sqrt{1+{a(\eta)\over a(\nq)}},
\label{xdefinition}
\end{equation}
so that
\begin{equation}
{dx\over d\eta}={1\over 2x}{\dot a(\eta)\over a(\nq)}.
\label{dxdn}
\end{equation}
Using (\ref{xdefinition}) and (\ref{dxdn}) one may
write the wave equation (\ref{wavetwo})
in the form
\begin{eqnarray}
{d^2\chi\over d x^2}&+&\bigg[2x {\ddot a(\eta)a(\nq)\over
{\dot a}^2(\eta)}-{1\over x}+2x{a(\nq)\over a(\eta)}\bigg]{d \chi
\over dx}\nonumber\\
&&\nonumber\\
&+&\bigg[4 k^2 x^2 {a^2(\nq)\over {\dot a}^2(\eta)}
-x^2 {a^2(\nq)\over a^2(\eta)}-2x^2 {\ddot a(\eta)a^2(\nq)\over
{\dot a}^2(\eta) a(\eta)}\bigg]\chi=0.
\label{wavethree}
\end{eqnarray}
This expression can be simplified 
using the Einstein equations (\ref{Einstein}), 
along with (\ref{mixedenergydensity})
and (\ref{xdefinition})
One can show that
\begin{equation}
{\dot a}^2(\eta) =2 x^2 \ddot a(\eta) a(\nq),
\end{equation}
and using this expression one may write the wave equation
(\ref{wavethree}) as
\begin{equation}
{d^2\chi\over d x^2}+2x{a(\nq)\over a(\eta)}{d\chi
\over d x}+\bigg[2k^2{a(\nq)\over \ddot a(\eta)} -x^2 {a^2(\nq)\over
a^2(\eta)}-{a(\nq)\over a(\eta)}\bigg]\chi=0.
\end{equation}
Since from (\ref{xdefinition})
\begin{equation}
{a(\nq)\over a(\eta)}={1\over x^2-1},
\end{equation}
and from (\ref{mixedscalefactor}) and (\ref{xidefinition})
\begin{equation}
{\ddot a(\eta)\over a(\nq)}={1\over 2\nend^2} 
\bigg({\xi^2\over 1+\xi}\bigg),
\end{equation}
the wave equation becomes 
\begin{equation}
{d^2\chi\over dx^2}-{2x\over (1-x^2)}{d\chi\over dx}
+\bigg\{{2\over (1-x^2)}+4\kappa
-{1\over (1-x^2)^2}\bigg\}\chi=0.
\label{diffeq2}
\end{equation}
Here $\kappa$ is defined by
\begin{equation}
\kappa\equiv\bigg({1+\xi\over \xi^2}\bigg) k^2\nend^2=
{4\pi^2\over \lambda_0^2 H_0^2}{(2+\zeq)\over (1+\zeq)^2}
\approx {4\pi^2\over \lambda_0^2 H_0^2 \zeq},
\label{kappadefinition}
\end{equation}
where $\lambda_0$ is the {\it present} day wavelength of the mode
with wavenumber $k$.
For the multipole moments of interest, $l\in [2,1000]$,
the contributions typically come from wavelengths in the
range $\kappa \in [10^{-3},10^3]$.
This form of the wave equation is the spheroidal
wave function differential equation (\ref{diffeq}), which
is discussed in detail in appendix \ref{appendixa}.

The solutions to the wave equation (\ref{kge})
during the mixed phase can be expressed in terms of
spheroidal wave functions.
By inspection of 
(\ref{xdefinition}) one sees that $x > 1$,
and so the possible solutions to the wave equation (\ref{diffeq2})
may be expressed as sums of any pair of
\begin{equation}
\chi(x,\kappa)={^j \S^1_2}(x,\kappa),
\>\>j=1,2,3,4,
\label{chixk}
\end{equation}
where ${^j \S^1_2}(x,\kappa)$ is a spheroidal wave
function.
Using (\ref{chidefinition}), (\ref{chixk}),
the normalization condition (\ref{phinormalization}), 
and the Wronskian relation
(\ref{wronskian}), one obtains a positive-frequency
mode function during the mixed phase
\begin{equation}
\phi^{(+)}_{\rm mix}(\eta,k)=
-2i\sqrt{{8\over 3\pi}\ed {a(\nend)\over a(\eta)}\bigg(
{1+\xi\over\xi}\bigg)}\>\nend^2 k^{1/2}
\>\>{^4 \S^1_2}(x,\kappa).
\label{mixedmodefunction}
\end{equation}
Note that $x$ is a function of the conformal time
$\eta$ and $\kappa$ depends on the wavenumber $k$.
The negative-frequency mode function $\phi_{\rm mix}^{(-)}(\eta,k)
= \left[\phi_{\rm mix}^{(+)}(\eta,k)\right]^*$
is just the complex
conjugate of the positive-frequency mode function.
The positive 
and negative frequency mode functions form a complete solution
to the wave equation during the mixed radiation
and matter phase. Note that the choice of graviton mode
function during the de Sitter phase completely determines
the mode function at all later times. Thus, the choice (\ref{mixedmodefunction})
of ``positive-frequency'' during the mixed phase is 
unimportant. 

\subsection{The Graviton Mode Function Expressed Using 
Bogolubov Coefficient Notation}

The choice of mode function during the de Sitter phase
completely determines the mode function at all later
times. This is because a solution to the wave
equation depends only on the values
of $\phi$ and $\dot\phi$ on a spacelike Cauchy surface
(ie. a surface of constant $\eta$). We choose the mode
function during the de Sitter phase to be the pure 
positive-frequency de Sitter solution (\ref{phidesitter}).
This is the unique solution corresponding to a de Sitter-invariant
vacuum state with the same (Hadamard) short distance behavior
as one would find in Minkowski space \cite{Allen}.
Having made this choice for the mode function during
the de Sitter phase, the mode function during all
times  is
\begin{equation}
\phi(\eta,k)=\left\{
\begin{array}{lll}
\phi^{(+)}_{\rm ds}(\eta,k)
 & \eta\leq\nend, &\text{de Sitter phase},\\ 
&&\\
\alpha(k\nend)\phi^{(+)}_{\rm mix}(\eta,k)
+\beta(k\nend)\phi^{(-)}_{\rm mix}(\eta,k)
& \eta\geq\nend,&\text{mixed phase},
\end{array}
\right.  
\label{finalmodefunction}
\end{equation} 
where $\alpha$ and $\beta$ are Bogolubov coefficients.

The Bogolubov coefficients are determined by requiring that the mode
function and its first derivative be continuous
at $\eta=\nend$.
One obtains
\begin{eqnarray}
\alpha(k\nend)&=&k\nend\sqrt{ {1+\xi\over\xi}}
\bigg\{\bigg[{\xi\over \sqrt{1+\xi}} 
{^3 \S^1_2}'(x_1,\kappa)-
{^3 \S^1_2}(x_1,\kappa)\bigg]
{i\over 2k\nend}\bigg(1+{i\over k\nend}\bigg)
-{^3 \S^1_2}(x_1,\kappa)\bigg\},\nonumber\\
&&\nonumber\\
\beta(k\nend)&=&k\nend\sqrt{ {1+\xi\over\xi}}
\bigg\{\bigg[{\xi\over \sqrt{1+\xi}} 
{^4 \S^1_2}'(x_1,\kappa)-
{^4 \S^1_2}(x_1,\kappa)\bigg]
{i\over 2k\nend}\bigg(1+{i\over k\nend}\bigg)
-{^4 \S^1_2}(x_1,\kappa)\bigg\}.\nonumber\\
&&
\label{bogolubov}
\end{eqnarray}
The prime in (\ref{bogolubov}) is defined by
\begin{equation}
{^j \S^\mu_\lambda}'(x_1,\theta)\equiv
\bigg[{\partial\over\partial z} {^j \S^\mu_\lambda}
(z,\theta)\bigg]_{z=x_1},
\end{equation}
where
\begin{equation}
x_1\equiv\sqrt{1+{a(\nend)\over a(\nq)}}=\sqrt{1+\xi}.
\end{equation}
Using the Wronskian relation for the spheroidal wave
functions (\ref{wronskian})  one can easily verify that
\begin{equation}
|\alpha(k\nend)|^2-|\beta(k\nend)|^2=1.
\end{equation}
We stress again that {\it the choice of graviton mode
function during the de Sitter phase completely determines
the mode function at all later times}. Thus, the choice
of ``positive-frequency'' during the mixed phase is 
unimportant. Had we chosen a different solution to
the wave equation during the mixed phase to
call ``positive-frequency'', then the Bogolubov coefficients
(\ref{bogolubov}) would be different in such a way so
that the mode function (\ref{finalmodefunction}) 
{\it would be the same}.


\section{Multipole Moments}
\label{multipolemoments}

Having determined the graviton mode function for our
cosmological model, we may calculate the
angular correlation function $C(\gamma)$, or equivalently, the
multipole moments $\langle a_l^2\rangle$. We simply substitute
the graviton mode function (\ref{finalmodefunction})
into the formulae (\ref{correlationdefinition}-\ref{Fdefinition}).

\subsection{Analytical Results}
\label{analyticalresults}

The graviton mode function (\ref{finalmodefunction}) is exact; no
approximations (ie. long wavelength) have been made.  One may directly
substitute the mode function into the formulae
(\ref{correlationdefinition}-\ref{Fdefinition}) to obtain an exact
expression for the multipole moments; because the arguments of the
spheroidal wave functions in (\ref{finalmodefunction}) are functions
themselves, however, the result is complicated and not very
illuminating. In typical inflationary models, the amount of expansion
is very large. If one takes the limit $\zend\rightarrow\infty$, then
$\xi\rightarrow 0$. This allows one to write a fairly compact
expression for the multipole moments; after substituting the mode
function (\ref{finalmodefunction}) into the formulae
(\ref{correlationdefinition}-\ref{Fdefinition}), one may collect
together terms in the expression for the multipole moments $\langle
a_l^2\rangle $ which are the same order in $\xi$. Then in the limit as
$\zend\rightarrow\infty$ and $\xi\rightarrow 0$, one may consider only
the leading term. The leading term, which is ${\rm O}(\xi^0)$,
is given below in (\ref{moments}). The neglected terms are 
${\rm O}(\xi^1)$ or greater.  For typical inflationary
models one has $\zend\gtrsim 10^{20}$ and $\zeq\approx 10^4$, so
that $\xi\lesssim 10^{-16}$. So the expression (\ref{moments}) is quite
accurate for most cosmological models, since the neglected terms are
very small.

The expressions for the multipole moments
$\langle a_l^2\rangle$ are fairly simple.
After substituting the mode function (\ref{finalmodefunction})
into the formulae (\ref{correlationdefinition}-\ref{Fdefinition}),
and making the same changes of variable as in section \ref{mixed}, ie.
\begin{equation}
x\equiv \sqrt{1+{a(\eta)\over a(\nq)}}\>\>\> \text{and}
\>\>\>\kappa\equiv\bigg( {1+\xi\over\xi^2}\bigg)k^2\nend^2,
\end{equation}
one obtains for the multipole moments $(l\geq 2)$
\begin{equation}
\al={4\over 3}\pi^2 {(l+2)!\over (l-2)!}\ed
\int_0^\infty {d\kappa\over \kappa^{5/2}}
\bigg\{{^1 \T^1_2}(x_1,\kappa)
G_l^2 (\kappa)-{^2 \T^1_2}(x_1,\kappa)
G_l^1 (\kappa)\bigg\}^2+{\rm O}(\xi^1).
\label{moments}
\end{equation}
The integral above is simply an integral over
the (rescaled, dimensionless) wavenumber $\kappa$.
The functions ${^j \T^1_2}(x_1,\kappa)$
are the same functions defined in (\ref{newTdefinition}).
The functions $G^j_l(\kappa)$ are the (reparametrized) 
Sachs-Wolfe integrals over null geodesics, given by
\begin{equation}
G_l^ j(\kappa)\equiv\int_{x_{\rm ls}}^{x_o} 
dx{J_{l+1/2}\Big(2\kappa^{1/2}(x_o-x)\Big)\over (x_o-x)^{5/2}
(x^2-1)^{1/2}}\bigg\{{x\over x^2-1}{^j \S^1_2}(x,\kappa)-
{^j \S^1_2}'(x,\kappa)\bigg\},
\label{Gldefinition}
\end{equation}
where the limits $x_{\rm ls}$ and $x_o$ on the integral are
given in terms of redshift by
\begin{equation}
x_o\equiv \sqrt{2+Z_{\rm eq}}\>\>\text{and}\>\>
x_{\rm ls}\equiv\sqrt{1+{1+Z_{\rm eq}\over 1+Z_{\rm  ls}}}.
\end{equation}
These formulae, and especially the spheroidal wave functions,
may be numerically evaluated using the techniques
discussed in appendix \ref{appendixb}. Henceforth
we refer to  (\ref{moments}) and
(\ref{Gldefinition}) as the ``mixed'' formulae for the multipole
moments. Furthermore, we refer to multipole
moments calculated using (\ref{moments}) and
(\ref{Gldefinition}) as the ``mixed'' multipoles.

\subsection{Numerical Results}
\label{numericalresults}

We have numerically evaluated the ``mixed" formulae
(\ref{moments}) and (\ref{Gldefinition}).  The second column of Table
\ref{table2} shows the ``mixed" multipoles.  The third column of Table \ref{table2}
shows the results obtained by evaluating Eq. (6.2) of Allen and Koranda
\cite{AllenKoranda}, which we refer to as the ``exact-dust" formula for
the multipole moments.  The ``exact-dust" formula assumes 
(1) the universe begins with an initial de Sitter phase, followed by
(2) a pure radiation phase containing only radiation and no dust,
followed by (3) a pure dust phase containing only dust and no
radiation, during which (4) last-scattering takes place. 
We refer to this type of universe  as a
``dust" universe, to distinguish it from the model described by
(\ref{mixedscalefactor}), (\ref{mixedenergydensity}), and
(\ref{mixedpressure}), which we refer to as ``mixed" universe
because it contains both dust and radiation at last-scattering.  The
fourth column shows the results obtained by evaluating Eq. (6.1) of
Allen and Koranda, which we refer to as the ``approximate-dust" formula for
the multipole moments.  The ``approximate-dust" formula is a
long-wavelength approximation to the ``exact-dust" formula, and is {\it
equivalent} to the standard formulae for the multipole moments given by
Abbott and Wise \cite{AbbotWise}  and Starobinsky \cite{Starobinsky}.
The fifth column shows recent results of Ng and Speliotopoulos
\cite{Ng}, who numerically integrated the wave equation to obtain the
amplitude of the gravitational wave (or graviton mode function). They
also considered a universe model which  contains both radiation and
dust at last-scattering.  All of the results in Table \ref{table2} were
obtained with $Z_{\rm eq}=6000$ and $Z_{\rm ls}=1100$.

Figures \ref{FIGURE1},\ref{FIGURE2}, and \ref{FIGURE3} compare the
``mixed" multipole moments 
to the multipole moments calculated using other techniques.
The quantity $M_l$ in Figures \ref{FIGURE1}-\ref{FIGURE3}
is defined by
\begin{equation}
M_l\equiv 
{l(l+1)\over 6}{\langle a_l^2\rangle\over \langle a^2_2\rangle} .  
\label{Mldefinition}
\end{equation}
(Note that in \cite{AllenKoranda} the equation
defining $M_l$ contains an extraneous factor of
$\rho_{\rm ds}/\rho_{\rm P}$.)
The points labeled ``mixed'' in Figure \ref{FIGURE1} show
results obtained by
numerically evaluating the ``mixed" formulae (\ref{moments}) and
(\ref{Gldefinition}). The  points in Figure \ref{FIGURE1} labeled ``transfer function''
show results from Turner, White, and Lidsey \cite{Turner2},
who numerically integrate the wave
equation, and express the solutions in terms of the
standard long-wavelength approximate mode functions \cite{AllenKoranda}
using a ``transfer function''. The points in Figure \ref{FIGURE2}
labeled ``mixed''
again show results obtained by numerically evaluating the ``mixed"
formulae.  
The points labeled ``exact-dust'' show
results obtained using the ``exact-dust" formula, and
the points labeled ``approximate-dust'' show results obtained 
using the ``approximate-dust" formula. Figure \ref{FIGURE3} compares the
``mixed" results to results from  Dodelson, Knox, and Kolb
\cite{KnoxDodelson} (labeled ``Boltzmann''), who do not use the Sachs-Wolfe
formula to calculate the CBR anisotropy.  Instead they use numerical
methods to evolve the photon distribution function using first-order 
perturbation theory of the general relativistic Boltzmann equation for
radiative transfer. All the multipole moments shown in Figures \ref{FIGURE1},
\ref{FIGURE2}, and \ref{FIGURE3} are for cosmological parameters
$\zeq\approx 6000$ and $Z_{\rm ls}=1100$. (The  values of
$\zeq$ for the Turner, White, and Lidsey work
and for the the Dodelson, Knox, and Kolb results 
differs from $6000$ by about one percent).

\subsection{Discussion}

The Ng and Speliotopoulos multipoles agree quite well with the ``mixed"
multipoles obtained here, using spheriodal wavefunctions, for a
cosmology containing dust and radiation components. This is expected
because the two methods used to calculate the multipole moments should
be essentially equivalent. The graviton mode functions Ng and
Speliotopoulos obtain by numerically integrating the wave equation
(with the correct boundary conditions) must be equivalent to our
analytic expressions (\ref{finalmodefunction}).  The cosmological model
Ng and Speliotopoulos consider, however, is slightly different than our
own. They model the smooth transition during the mixed phase by a
simpler scale factor than our own (\ref{mixedscalefactor}).  This may
account for the small discrepancy between their results and our own.

The Dodelson, Knox, and Kolb  (Figure \ref{FIGURE3}) multipoles also
appear consistent with the ``mixed" multipoles.  This is also expected
since the Boltzmann formalism \cite{Crittenden,KnoxDodelson} and the
Sachs-Wolfe formalism should yield equivalent results.  We do not
understand at this time the small discrepancy between the two multipole
spectrums.

For $l\lesssim 30$, the ``exact-dust" multipoles and the
``approximate-dust" multipoles agree fairly well with the ``mixed"
multipoles (Figure \ref{FIGURE2}).  This is not surprising. The small
$l$ multipoles $\langle a_l^2\rangle$ are most affected by longer
wavelength perturbations. These longer wavelength perturbations were
redshifted outside the Hubble sphere early in the inflationary phase,
and only recently re-entered the Hubble sphere (the longest wavelength
perturbations  remain outside the Hubble sphere even today
\cite{KolbTurner}). Because they remained outside the Hubble sphere
until recently, the longer wavelength perturbations are insensitive to
the details of the evolution of the universe before it became dust
dominated. Thus, the ``mixed" universe and ``dust" universe models are
essentially equivalent for longer wavelength perturbations.  So it's
not surprising that the same CBR anisotropies are produced by the long
wavelength perturbations in either model. Thus
the ``exact-dust", ``approximate-dust", and ``mixed" multipoles should,
and do, agree for small $l$.  Furthermore, because the
``approximate-dust" formula for the multipole moments is equivalent to
the standard formulae given by Abbott and Wise \cite{AbbotWise} and
Starobinsky \cite{Starobinsky}, one can conclude that the standard
formulae for the multipole moments are accurate for small $l$, whether
or not the universe was completely dust dominated at last-scattering.

For $l\gtrsim 30$, the ``exact-dust" multipoles and the
``approximate-dust" multipoles differ significantly from the ``mixed"
multipoles.  Again this is not surprising. The larger $l$ multipoles
are more affected by shorter wavelength perturbations, which re-entered
the Hubble sphere before the universe became dust dominated, and are
therefore sensitive to the details of the cosmological expansion before
dust domination.  A ``mixed" universe becomes dust dominated {\it much
more slowly} than a ``dust" universe, which is dust dominated
immediately after the radiation phase ends at dust/radiation equality
$\nq$. Therefore the shorter wavelength modes in a ``mixed" universe
and  a ``dust" universe evolve very differently after they re-enter the
Hubble sphere. This difference is evident in Figure \ref{FIGURE2} for
the large $l$ multipole moments.

The  Turner, White, and Lidsey (TWL) multipole spectrum differs
significantly from the ``mixed" multipole spectrum (Figure
\ref{FIGURE1}).  In particular their multipole moments are
significantly greater than the ``mixed" multipoles for $3\lesssim
l\lesssim 80$.  Although TWL consider a ``mixed" universe, they use the
standard multipole moment formulae for a ``dust" universe. However,
they modify the standard formulae by including a time-independent
transfer function $T(k/k_{\rm eq})$, which depends only on the
wavenumber $k$.  TWL give an explicit functional form for the transfer
function, which they obtain by numerically integrating the wave
equation \cite{Turner2}:
\begin{equation}
T(y)=[1.0+1.34y+2.50y^2]^{1/2},\>\text{where}\>y=k/k_{\rm eq}.
\label{numericaltransferfunction}
\end{equation}
Since the transfer function $T \ge 1$ appears in the expression for the
multipole moments as $|T(k/k_{\rm eq})|^2$ (see Eqs. (22) and (23) of
\cite{Turner2}), one can see that the effect of the transfer function
(\ref{numericaltransferfunction}) is to {\it increase} the multipole
moments. Furthermore, the contribution from shorter wavelength (larger
$k$) modes is enhanced more than the contribution from larger
wavelength (smaller $k$) modes, since $T(k/k_{\rm eq})\rightarrow 1$ as
$k\rightarrow 0$.

Since the TWL multipoles are significantly greater than the ``mixed"
multipoles for $3\lesssim l\lesssim 80$, {\it the transfer function
overestimates the contribution to the multipole moments from longer
wavelength modes}. This is easy to see. As discussed above, the
standard formulae are accurate for small $l$ multipoles, and give
essentially the same results as the ``mixed" formulae for $l\lesssim 30$.
The TWL formulae for the multipole moments are equivalent to the
standard formulae, except for the transfer function. Thus if the
transfer function was set to unity, the TWL multipoles would be the
same as the standard multipoles, and hence would be equivalent to the
``mixed" multipoles for $l\lesssim 30$. Since the TWL multipoles are
larger than the ``mixed" multipoles for $3\lesssim l\lesssim 80$, the
transfer function must enhance the the contribution to the $3\lesssim
l\lesssim 80$ multipole moments too much. Because the small $l$ moments
are affected most by longer wavelength perturbations, the transfer
function must overestimate the contribution from longer wavelengths, or
smaller wavenumber $k$.

The TWL multipole spectrum  is also significantly different than the
``mixed" multipole spectrum for large $l$. The large $l$ multipoles are
most affected by shorter wavelength modes.  The transfer function
(\ref{numericaltransferfunction}) significantly enhances the
contribution of shorter wavelength modes (larger $k$) to the multipole
moments.  Because the TWL transfer function is time-independent,
however, it can not alter the time evolution of the shorter wavelength
modes. By comparing the ``dust" universe graviton mode function (see
Eq. (4.28) of \cite{AllenKoranda})
\begin{equation}
\phi(\eta,k)_{\rm dust}=\sqrt{{24\over \pi}{\rho_{\rm dS}\over \rho_{\rm P}}}
{j_1\Big(k(\eta+\nq)\Big)\over k^{5/2}(\eta+\nq)},
\end{equation}
(which is equivalent to the standard formulae for the gravity-wave
amplitude \cite{AllenKoranda}, and hence equivalent to the TWL formulae
for the gravity-wave amplitude) to the ``mixed" universe mode function
(\ref{finalmodefunction}), one can see that the time evolution of the
shorter wavelength modes is very different in a ``dust" universe when
compared to the time evolution in a ``mixed" universe. Thus one expects
that for shorter wavelength modes, the integral along the null
geodesics (\ref{Ildefinition}) in the Sachs-Wolfe formula will be very
different for a ``dust" universe, when compared to a ``mixed" universe.
Since the transfer function is time-independent, it has no effect on
the integral (\ref{Ildefinition}). So one expects that the multipole
spectrum for a ``dust" universe will be very different than the
multipole spectrum for a ``mixed" universe, even with the transfer
function included. Figure \ref{FIGURE1} shows that this is the case. In
particular, the ``bump'' in the TWL multipole spectrum appears near
$l\approx 180$, rather than $l\approx 200$, as for the ``mixed" multipole
spectrum.


\section{Conclusion}

This paper examines the tensor perturbations of the gravitational field
in a spatially flat, FRW cosmology containing a mixture of radiation
and dust, and shows that they may be expressed in terms of spheroidal
wave functions.  Although spheroidal wave functions have appeared in
this context before, previous authors incorrectly determined the
characteristic exponent which labels these functions.  After explaining
the correct method for determining the characteristic exponent, we show
that spheroidal wave functions may be efficiently and accurately
evaluated using standard numerical techniques.

We considered inflationary cosmological models, and used the spheroidal
wave functions to find the spectrum of CBR temperature fluctuations
resulting from primordial tensor (gravitational radiation)
perturbations.  These temperature fluctuations are predicted by all
inflationary models.  Their existence follows from first principles: it
is a consequence of the uncertainty principle and the Einstein
equation.  The temperature fluctuations have been previously studied by
a number of authors (including ourselves) using a variety of
approximations, and both analytic and numerical techniques.

In hindsight, only two approximations remain in this work.  One is that
the amplitude of the gravitational perturbation $h_{ij}$ is very
small.  This approximation is indeed well justified.  The second
approximation is that the energy density and pressure of the universe
correspond to a mixture of dust and radiation as given in 
(\ref{mixedenergydensity}-\ref{mixedpressure}).  Going back in time, this approximation is
good until approximately the time of nucleosynthesis, $t=200$ seconds,
when the number of effectively massless particles in the universe
changed.  This should not affect the multipole moments 
$\langle a_l^2\rangle$ which we consider, which have 
$l<300 $.  For larger values of $l$, other physical effects such as the
finite thickness of the surface of last scattering may also become
relevant.

It is  likely that the spheroidal wave functions, which describe
gravitational wave perturbations in realistic cosmological models, will
find other useful applications.  We expect that the results and methods
of this paper may  prove applicable to a wider variety of
calculations than the CBR temperatures perturbations considered here.

\acknowledgements
We are grateful to Scott Dodelson, Edward Kolb, Lloyd Knox, Jorma Louko, and Mike Turner
for useful comments and suggestions. 
This work has been partially supported by NSF Grant No.
PHY91-05935.


\appendix

\section{}
\label{appendixa}

In spatially flat FRW universes, graviton mode functions in a
linearized theory of gravity obey a minimally-coupled, massless, scalar
wave equation \cite{FordParker} like (\ref{kge}).  If the scale factor
of the spatially flat FRW universe transforms smoothly from being
radiation  to  dust dominated, the solutions to the equation are
spheroidal wave functions.

\subsection{The  Differential Equation of Spheroidal Wave Functions
and Its Solutions}

There is no generally accepted form for the differential equation of
spheroidal wave functions. We write the differential equation as
\begin{equation}
{d ^2\varphi\over d z^2}-{2z\over (1-z^2)}
{d \varphi\over d z}
+\bigg\{{\lambda\over (1-z^2)}+4\theta-{\mu^2\over (1-z^2)^2}\bigg\}\varphi=0.
\label{diffeq}
\end{equation}
The parameters $\mu,\lambda$, and $\theta$  and the variable $z$  can
in general be complex.  Here we take $z$,$\lambda,\mu$, and $\theta$ to
be real.  We also consider only $\theta\geq 0$.
The differential equation (\ref{diffeq}) has two regular
singular points at $z=\pm1$ and an irregular singular point at
$z=\infty$.  We only consider $z>1$. Then the solutions of
(\ref{diffeq}) are the {\it spheroidal wave functions}
\cite{Erdelyi,MeixnerSchafke}
\begin{equation}
\varphi=S^{\mu (j)}_\nu (z,\theta),\>\> z>1,\>\>j=1,2,3,4.
\label{solutions}
\end{equation}
The parameter $\mu$ is the {\it order} of the spheroidal wave function,
and $\nu$ is the {\it characteristic exponent}. In later sections we
consider in detail the characteristic exponent $\nu$ and its relation
to the order $\mu$ and the parameters $\lambda$ and $\theta$. For now
we simply note that $\nu$ is restricted so that
\begin{equation}
\nu+1/2\neq\>{\rm integer}.
\label{nohalfinteger}
\end{equation}
For a very thorough and complete discussion of the differential
equation (\ref{diffeq}) and the solutions (\ref{solutions}) see
reference \cite{MeixnerSchafke}.

The spheroidal wave functions can be expressed in terms of
more familiar special functions.  If $\theta=0$ then the
differential equation (\ref{diffeq}) reduces to Legendre's differential
equation, suggesting that the spheroidal wave functions can be
expressed in terms of Legendre functions
\cite{Erdelyi,MeixnerSchafke}.  For $z>1$, however, it is more useful
to express the spheroidal wave functions as infinite sums of Bessel functions.
For $\mu>0$ one can write the spheroidal functions as
\begin{equation}
S^{\mu (j)}_\nu (z,\theta)= \bigg({z^2\over z^2-1}\bigg)^{\mu/2}
T^{\mu(j)}_\nu (z,\theta),\>\>\mu>0,
\label{sdefinition}
\end{equation}
where $T^{\mu(j)}_\nu (z,\theta)$ is the infinite sum
\begin{equation}
T^{\mu(j)}_\nu (z,\theta)=s^\mu_\nu(\theta)\sum_{r=-\infty}^\infty
a^\mu_{\nu,r}(\theta)\psi^{(j)}_{\nu+2r}(2\theta^{1/2}z),\>\>
j=1,2,3,4.
\label{tdefinition}
\end{equation}
The expansion coefficients
$a^\mu_{\nu,r}(\theta)$ and the normalization factor
$s^\mu_\nu(\theta)$, which are the same for any $j$,
are discussed below.
The functions $\psi^{(j)}_\upsilon(z)$ are
proportional to Bessel or Hankel functions:
\begin{eqnarray}
\psi_\upsilon^{(1)}(z)&=\sqrt{{\pi\over 2z}}
 J_{\upsilon+1/2}(z),\nonumber\\
\psi_\upsilon^{(2)}(z)&=\sqrt{{\pi\over 2z}}
 Y_{\upsilon+1/2}(z),\nonumber\\
\psi_\upsilon^{(3)}(z)&=\sqrt{{\pi\over 2z}}
 H^{(1)}_{\upsilon+1/2}(z),\nonumber\\
\psi_\upsilon^{(4)}(z)&=\sqrt{{\pi\over 2z}}
 H^{(2)}_{\upsilon+1/2}(z).
\label{psidefinition}
\end{eqnarray}
The spheroidal wave function of the first kind $(j=1)$
and the spheroidal wave function of the second kind $(j=2)$ 
form a complete solution to the differential equation (\ref{diffeq}).
Because the Hankel functions 
$H^{({1\atop 2})}_\upsilon(z)=J_\upsilon(z)\pm i Y_\upsilon(z)$
can be written as linear combinations of
Bessel functions (see equation (3.86) in reference \cite{Jackson}), 
the spheroidal functions of
the third $(j=3)$ and fourth $(j=4)$ kind can be written as linear
combinations of spheroidal functions of the first and second kind:
\begin{eqnarray}
S^{\mu (3)}_\nu (z,\theta)&=&S^{\mu (1)}_\nu (z,\theta)+
iS^{\mu (2)}_\nu (z,\theta),\nonumber\\
S^{\mu (4)}_\nu (z,\theta)&=&S^{\mu (1)}_\nu (z,\theta)-
iS^{\mu (2)}_\nu (z,\theta). 
\label{s3s4definition}
\end{eqnarray}
The spheroidal functions of the third  and fourth  kind  also form a
complete solution to the spheroidal differential equation.  The
expansions (\ref{sdefinition}) of the spheroidal wave functions in
terms of Bessel and Hankel functions are only useful if the infinite
sums (\ref{tdefinition}) converge.

The convergence of the infinite sums  depends on the 
expansion coefficients $a^\mu_{\nu,r}(
\theta)$. Substituting (\ref{sdefinition}) and (\ref{tdefinition})
into the differential equation (\ref{diffeq}) yields a three-term recurrence
relation that the expansion coefficients must satisfy. 
The recurrence relation can be written as
\begin{equation}
\A^\mu_{\nu,r}(\theta) a^\mu_{\nu,r-1}(\theta) +
\B^\mu_{\nu,r}(\theta) a^\mu_{\nu,r}(\theta) +
\C^\mu_{\nu,r}(\theta) a^\mu_{\nu,r+1}(\theta)=0,
\label{recurrencerelation}
\end{equation}
where 
\begin{eqnarray}
\A^\mu_{\nu,r}(\theta)&=&
4\theta{(\nu+2r-\mu)(\nu+2r-\mu-1)\over (2\nu+4r-3)(2\nu+4r-1)},
\nonumber\\
&&\nonumber\\
\B^\mu_{\nu,r}(\theta)&=&
\lambda-(\nu+2r)(\nu+2r+1)+{(\nu+2r)(\nu+2r+1)+\mu^2-1\over
(2\nu+4r-1)(2\nu+4r+3)}8\theta,\label{ABCdefinition}
\\
&&\nonumber\\
\C^\mu_{\nu,r}(\theta)&=&
4\theta{(\nu+2r+\mu+2)(\nu+2r+\mu+1)\over (2\nu+4r+3)(2\nu+4r+5)}.
\nonumber
\end{eqnarray}
The solution to this recurrence relation, as well as the convergence of
the infinite sums (\ref{tdefinition}), depends critically on the
parameters $\mu,\nu,\lambda$, and $\theta$, and is discussed below.  
Once a solution to the recurrence
relation (for which the infinite sums (\ref{tdefinition}) converge)
is obtained, the normalization factor $s^\mu_\nu(\theta)$ is
given by
\begin{equation}
s^\mu_\nu (\theta)=\big[\sum_{r=-\infty}^{\infty} (-1)^r a^\mu_{\nu,r}
(\theta)\big]^{-1}.
\label{snormalization}
\end{equation}
This normalization is chosen so that in the limit as $z$
becomes very large
\begin{equation}
\lim_{z\rightarrow \infty}\Big[ S^{\mu(j)}_\nu (z,\theta)
/\psi_\nu^{(j)}(2\theta^{1/2} z)\Big]=1.
\label{latetime}
\end{equation}
This relation and many more details of the solutions
(\ref{sdefinition}) are developed in reference \cite{MeixnerSchafke}.

\subsection{The Eigenvalue \protect{$\lambda$} }

Although the order $\mu$ of the spheroidal wave function, along with
the parameters $\theta$ and $\lambda$, appears directly in the
differential equation (\ref{diffeq}), the characteristic exponent $\nu$
does not. In most investigations and applications of spheroidal wave
functions
\cite{Erdelyi,MeixnerSchafke,AbramowitzStegun,SpheroidalStudies},
however, the parameter $\lambda$ is left unfixed; one {\it assumes} a
(typically integer) value for $\nu$ and considers $\lambda$ to be a function
of $\mu,\nu$, and $\theta$ and writes 
\begin{equation}
\lambda=\lambda^\mu_\nu (\theta).
\end{equation}
$\lambda^\mu_\nu (\theta)$ is often referred to as an eigenvalue,
especially when considering spheroidal wave functions as
solutions to the three-dimensional wave equation \cite{MorseFeshbach}.

For a given choice of the parameters $\mu,\nu$, and $\theta$, the
eigenvalue $\lambda^\mu_\nu (\theta)$ is that value of $\lambda$ for
which the recurrence relation (\ref{recurrencerelation}) has a {\it
minimal} solution. A minimal solution, roughly speaking, is a set of
coefficients $a^\mu_{\nu,r}(\theta)$ that satisfy the recurrence
relation {\it and} fall off for large $|r|$ \cite{Gautschi}.  A {\it
dominant} solution is a set of coefficients that satisfy the recurrence
relation but do not fall off. If the solution to the recurrence
relation is a dominant solution, the coefficients
$a^\mu_{\nu,r}(\theta)$ do not fall off for large $|r|$, and the
infinite sums in (\ref{tdefinition}) may not converge.  In general a
three-term recurrence relation will have two independent solutions,
much like a second-order, ordinary differential equation. However,
neither of the two solutions, nor any linear combination of the two
solutions, need be a minimal solution \cite{Gautschi}.  For a given set
of parameters $\mu,\nu$, and $\theta$, a minimal solution to the
recurrence relation (\ref{recurrencerelation}) exists only for a
single, discrete value of $\lambda$. That value for which the minimal
solution exists is the eigenvalue $\lambda^\mu_\nu (\theta)$.  In our
problem, we are given $\lambda,\theta$, and $\mu$. One can find $\nu$
(modulo an integer) by requiring that the recurrence relation
(\ref{recurrencerelation}) have a minimal solution.

The functional relationship between the parameters $\mu,\nu,\theta$ and
the eigenvalue $\lambda^\mu_\nu(\theta)$ is complicated.
It has been shown \cite{Erdelyi} that the functional relationship can
be expressed as
\begin{equation}
\cos(2\pi\nu)=f(\lambda,\mu^2,\theta),
\label{constraint}
\end{equation}
where a closed, analytic form for the function $f$ is usually
unattainable. The relation (\ref{constraint}) only determines
the characteristic exponent $\nu$ (as a function of 
$\lambda,\mu$, and $\theta$) up to an integer; 
a second constraint \cite{MeixnerSchafke}
fixes $\nu$:
\begin{equation}
\lambda^\mu_\nu(\theta=0)=\nu(\nu+1).
\label{constraint2}
\end{equation}
In section \ref{mixed} the special case of the differential
equation (\ref{diffeq}) with $\lambda=2$ is considered. The constraint
(\ref{constraint2}), along with the condition (\ref{nohalfinteger})
that $\nu$ not be a half-integer, fixes $\nu$ so that
\begin{equation}
{1\over 2}<\nu<{3\over 2} \>\>\text{for}\>\> \lambda=2.
\end{equation}
For investigations of some
of the analytic properties of $f(\lambda,\mu^2,\theta)$
see reference \cite{MeixnerSchafke}.
A more practical method for finding the functional relation between the
parameters $\mu,\nu,\theta$, and $\lambda$ is discussed in
section \ref{characteristicexponent}.

\subsection{The Case \protect{$\mu=1$}, \protect{$\lambda=2$}}

The special case of the differential equation (\ref{diffeq})
with $\mu=1$ and $\lambda=2$ is of cosmological interest,
and has been previously studied by Sahni \cite{Sahni} and Nariai \cite{Nairi},
who incorrectly take the solution to be $S^{1(4)}_1$.
In  spatially flat FRW universes containing a mixture of dust and radiation,
one may cast the wave equation for graviton mode functions in the form of
the spheroidal wave function differential equation, as shown in
section \ref{mixed}.  The differential equation  the graviton
mode function obeys (\ref{diffeq2}) is 
the special case of the differential equation
(\ref{diffeq}) with $\mu=1$, $\lambda=2$, and $\theta$ arbitrary.
($\theta$ is arbitrary
since in section \ref{mixed} it is proportional to the wave
number of a graviton mode function, and we desire expressions for the
mode functions that are valid for an interesting range of wavenumbers.)
Even if one   takes $\lambda$ in (\ref{diffeq}) to be fixed and does not
consider the eigenvalue problem for $\lambda$, the arguments and
conclusion in the previous paragraph are still valid, especially the constraint
(\ref{constraint}); if $\mu,\nu,\lambda,$ and $\theta$ do not
satisfy (\ref{constraint}) a minimal solution to the recurrence
relation does not exist and the infinite sums in (\ref{tdefinition})
do not converge. If $\lambda=2$, $\mu=1$, and $\theta$
is arbitrary, then the characteristic exponent $\nu$  is determined
by the constraint (\ref{constraint}) with $\lambda=2$ and $\mu=1$.
Tables  of eigenvalues
$\lambda^\mu_\nu (\theta)$ for different values of $\mu,\nu$, and
$\theta$ have been published\cite{SpheroidalStudies}, however, and from these
one can determine that the solution to (\ref{constraint}) for $\mu=1$,
$\lambda=2$, and $\theta$ arbitrary is {\it not} $\nu=1$, ie.
\begin{equation}
\cos(2\pi)\neq f(2,1,\theta)\>\>\text{for arbitrary}\>\>\theta. 
\end{equation}
So for arbitrary $\theta$ the solution
to the differential equation (\ref{diffeq}) with 
$\mu=1$ and $\lambda=2$ is {\it not}
the spheroidal wave function (\ref{solutions}) with
$\mu=1$ and $\nu=1$, ie.
\begin{equation}
\varphi\neq S^{1(j)}_1(z,\theta)\>\>\text{for}\>\>\lambda=2.
\end{equation}
This subtle point is missed in \cite{Sahni,Nairi}, where the 
solution to the spheroidal wave function differential equation
with $\mu=1$, $\lambda=2,$ and $\theta$ arbitrary is incorrectly given
as $S^{1(4)}_1$.

\subsection{A More Suitable Notation}
\label{betternotation}
The  notation introduced above for the spheroidal wave functions
(\ref{solutions})  is most useful when the order $\mu$ and the
characteristic exponent $\nu$ are fixed, and the eigenvalue
$\lambda^\mu_\nu(\theta)$ is considered as a function of $\mu,\nu,$
and $\theta$.  Since we are most interested in the case when $\mu$ and
$\lambda$ are fixed, it is more descriptive to denote the characteristic
exponent as $\nu=\nu^\mu_\lambda(\theta)$.  This convention, however,
would have one denote the spheroidal wave functions as
$S^{\mu(j)}_{\nu^\mu_\lambda(\theta)} (z,\theta)$, which is cluttered
and inconvenient. In section \ref{mixed}, where $\theta$ is itself a
function (of wavenumber), the notation would be even more unpleasant.

For this reason we adopt a new notation for the spheroidal wave
functions. The solution to the differential equation (\ref{diffeq}) 
with $\mu$ and $\lambda$ fixed is denoted
\begin{equation}
{^j \S^\mu_\lambda}(z,\theta)\equiv 
S^{\mu(j)}_{\nu^\mu_\lambda(\theta)} (z,\theta).
\end{equation}
The function on the left-hand side above is, of course, the same
function as on the right-hand side,  expressed using a different
notation. The indices $\mu$ and $\lambda$ are those that appear in the
differential equation, and again $j=1,2,3,4$. In terms of Bessel
functions, the spheroidal wave function is now written as
\begin{equation}
{^j \S^\mu_\lambda}(z,\theta)=\bigg({z^2\over z^2-1}\bigg)^{\mu/2}\>
{^j \T^\mu_\lambda}(z,\theta),
\end{equation}
where
\begin{equation}
{^j \T^\mu_\lambda}(z,\theta)\equiv
s^\mu_\lambda(\theta)\sum_{r=-\infty}^\infty
a^\mu_{\lambda,r}(\theta)\psi^{(j)}_{\nu+2r}(2\theta^{1/2}z),
\label{newTdefinition}
\end{equation}
 with $\nu=\nu^\mu_\lambda(\theta)$, 
 and the $\psi^{(j)}_\upsilon$ are the same functions given in
(\ref{psidefinition}).
The expansion coefficients $a^\mu_\lambda(\theta)$ are the same
as those in (\ref{tdefinition}) and obey the same recurrence relation
(\ref{recurrencerelation}) with the obvious change in notation.
The normalization factor $s^\mu_\lambda(\theta)$ is also the same
as in (\ref{tdefinition}) with the obvious change in notation.
We use this notation in sections \ref{mixed}-\ref{analyticalresults},
and the rest of this appendix.

\subsection{Wronskian and Complex Conjugates}

Two relations for the spheroidal wave functions are especially
useful in section \ref{mixed}. The first involves the spheroidal
wave functions of the third and fourth kind, and is obvious
from the relations (\ref{s3s4definition}):
\begin{equation}
{^4 \S^\mu_\lambda}(z,\theta)=\big[{^3 \S^\mu_\lambda}(z,\theta)\big]^*.
\label{complexconjugate}
\end{equation}
Here a $*$ denotes complex conjugation. The second relation is
the Wronskian for the spheroidal wave functions of the
third and fourth kind:
\begin{equation}
W\big[{^4 \S^\mu_\lambda}(z,\theta),{^3 \S^\mu_\lambda}(z,\theta)]
\equiv {^4 \S^\mu_\lambda}(z,\theta){d\over dz}{^3 \S^\mu_\lambda}(z,\theta)-
{^3 \S^\mu_\lambda}(z,\theta){d\over dz}{^4 \S^\mu_\lambda}(z,\theta)
={i\over \theta^{1/2}(z^2-1)}.
\label{wronskian}
\end{equation}
This relation, along with related results, is derived in 
reference \cite{MeixnerSchafke} (see especially section 3.65, equation
(53) ).

\subsection{A Practical Method for Determining 
\protect{$\nu^\mu_\lambda(\theta)$}
 and \protect{${^j \S^\mu_\lambda}(z,\theta)$}}
\label{characteristicexponent}

The constraints (\ref{constraint}) and (\ref{constraint2})
determine $\nu^\mu_\lambda(\theta)$
for a given choice of $\mu,\lambda$, and $\theta$.
In practice, however, it is difficult to use these constraints since a
closed, analytic form for the the function $f$ is usually not known.
Recall, though, that the constraint (\ref{constraint}) is equivalent to finding a minimal
solution to the recurrence relation (\ref{recurrencerelation}).  A
minimal solution exists only for the single, discrete value of
$\nu^\mu_\lambda(\theta)$ that satisfies the constraints
(\ref{constraint}) and (\ref{constraint2}).

One can use the theory of continued fractions to find the minimal
solution to the recurrence relation and determine $\nu^\mu_\lambda(\theta)$.
Divide (\ref{recurrencerelation}) by $a_r(\theta)$ and
consider the infinite continued fraction $P^\mu_{\lambda,r}(\theta)$
defined for $r\geq 1$ as
\begin{equation}
P^\mu_{\lambda,r}(\theta)\equiv{a_r\over a_{r-1}}=
{-\A_r\over \B_r
+\C_r
{\displaystyle {a_{r+1}\over a_r}}} ={-\A_r\over \B_r
+\C_r\displaystyle{-\A_{r+1}\over \B_{r+1}+\C_{r+1}\displaystyle{-\A_{r+2}\over
\B_{r+2}+\ldots}}}
\>\>r\geq 1,
\label{fraction1}
\end{equation}
where we have again suppressed for clarity the indices $\mu$ and
$\lambda$ as well as the argument $\theta$.   For a given choice of
parameters $\mu,\lambda,\theta$, and $\nu$ the infinite continued
fraction  can be used to find the ratios $a_r/a_{r-1}$ of all the
expansion coefficients, for $r\geq 1$, {\it if the continued fraction
converges}. There is no loss of generality if one assumes $a_0=1$
\cite{Erdelyi,MeixnerSchafke}, (note that this is compensated for in
the normalization $s^\mu_\lambda (\theta)$) so the continued fraction
(if it converges) can be used to find the expansion coefficients $a_r$
for $r\geq 1$.

A second infinite continued fraction
for $r\leq -1$ can also be derived from the recurrence relation.
We define the continued fraction
\begin{equation}
N^\mu_{\lambda,r}(\theta)\equiv{a_r\over a_{r+1}}=
{-\C_r\over \B_r
+\A_r
{\displaystyle {a_{r-1}\over a_r}}} ={-\C_r\over \B_r
+\A_r\displaystyle{-\C_{r-1}\over \B_{r-1}+\A_{r-1}\displaystyle{-\C_{r-2}\over
\B_{r-2}+\ldots}}}
\>\>r\leq -1.
\label{fraction2}
\end{equation}
Assuming $a_0=1$, the continued fraction (if it converges) can be
used to find the expansion coefficients $a_r$ for $r\leq -1$.

Whether or not the infinite continued fractions converge depends on the
existence of minimal solutions to the recurrence relation.  
Pincherle's Theorem \cite{Gautschi} tells
us that the infinite continued fraction $P^\mu_\lambda(\theta)$
converges for $r\geq 1$ if and only if the recurrence relation
has a {\it minimal} solution for $r\geq 1$.
Further, if the infinite continued fraction does converge, it converges
to a minimal solution.  Likewise, $N^\mu_\lambda(\theta)$ converges
to a minimal solution for $r\leq -1$ if and only if the recurrence
relation has a minimal solution for $r\leq -1$.

A subtle, but important, point is that $P^\mu_\lambda(\theta)$ and
$N^\mu_\lambda(\theta)$ may converge {\it to different minimal
solutions}. Given a set of parameters $\mu,\lambda,\theta$ and $\nu$,
the continued fraction (\ref{fraction1}) (with $a_0=1$) may converge
and determine a set of coefficients $a_1, a_2,a_3,\ldots$.  These
coefficients will satisfy the recurrence relation for $r\geq 1$
and will fall off for large $r$. Likewise,
for the same set of parameters, the continued fraction
(\ref{fraction2}) may converge and determine a set of coefficients
$a_{-1},a_{-2},a_{-3},\ldots$, which will satisfy the recurrence
relation for $r\leq -1$ and will fall of for large $|r|$. The
recurrence relation for $r=0$, however, {\it may not be satisfied}.
This is because the $r=0$ recurrence relation
\begin{equation}
\A^\mu_{\lambda,0}(\theta)a^\mu_{\lambda,-1}(\theta)
+\B^\mu_{\lambda,0}(\theta)a^\mu_{\lambda,0}(\theta)
+\C^\mu_{\lambda,0}(\theta)a^\mu_{\lambda,1}(\theta)=0
\end{equation}
is not explicitly solved when calculating either
the $P^\mu_\lambda(\theta)$ or the
$N^\mu_\lambda(\theta)$. One can see this by examining the
continued fractions (\ref{fraction1}) and (\ref{fraction2});
the three coefficients $a_{-1},a_0$, and $a_1$ do not
appear together anywhere in (\ref{fraction1}) and (\ref{fraction2}) for any $r$.
So both the set of expansion coefficients $a_1, a_2,a_3,\ldots$
found using $P^\mu_\lambda(\theta)$ and the set $a_{-1},a_{-2},a_{-3},\ldots$
found using $N^\mu_\lambda(\theta)$  are minimal solutions
for a certain range of the index $r$, but not for all $r$.

In order for the recurrence relation
(\ref{recurrencerelation}) to be satisfied it must be true for {\it
all} $r$. To ensure this, one must match the two solutions found using
the continued fractions. 
This is accomplished by requiring that
\begin{equation}
\A^\mu_{\lambda,0}(\theta)N^\mu_{\lambda,-1}(\theta)
+\B^\mu_{\lambda,0}(\theta)
+\C^\mu_{\lambda,0}(\theta)P^\mu_{\lambda,1}(\theta)\equiv Z^\mu_\lambda
(\theta,\nu)=0,
\label{Zdefinition}
\end{equation}
which is just the requirement that the recurrence relation be satisfied
for $r=0$. So finding the minimal solution to the recurrence relation,
and hence the characteristic exponent $\nu^\mu_\lambda(\theta)$, is
equivalent to finding the root of the the function
$Z^\mu_\lambda(\theta,\nu)$ defined in (\ref{Zdefinition}).  Given a
test value for $\nu^\mu_\lambda(\theta)$, one calculates the continued
fractions $P^\mu_{\lambda,1}(\theta)$ and $N^\mu_{\lambda,-1}(\theta)$,
and the rational functions $\A^\mu_{\lambda,0},\B^\mu_{\lambda,0}$, and
$\C^\mu_{\lambda,0}$, to find $Z^\mu_\lambda(\theta,\nu)$.  If
$Z^\mu_\lambda$ is not zero, one modifies the test value for
$\nu^\mu_\lambda(\theta)$ using whichever root-finding algorithm one
prefers.

In practice, this method for determining $\nu^\mu_\lambda(\theta)$
is efficient. Figure \ref{FIGURE4} shows the function
$Z^1_2(\theta,\nu)$ (of interest for the cosmological case)
for $\theta=5$ and $\frac{1}{2}<\nu<\frac{3}{2}$.
This plot is typical for $\theta$ in the range
$10^{-4}\leq\theta\leq 10^4$;  the root $\nu^\mu_\lambda(\theta)$
is always located between two singularities, and $Z^1_2$ is positive
for $\nu=\nu^\mu_\lambda(\theta)-\epsilon$ and negative for
$\nu=\nu^\mu_\lambda(\theta)+\epsilon$ where $0<\epsilon\ll 1$.
This   assists
implementing a root-finding algorithm to find the zeroes
of $Z^1_2(\theta,\nu)$ for arbitrary $\theta$. Figure
\ref{FIGURE5}  shows the characteristic exponent
$\nu^1_2(\theta)$ for $10^{-3}\leq\theta\leq 10^3 $.

Once the characteristic exponent $\nu^\mu_\lambda(\theta)$
is determined, one can 
use the continued fractions (\ref{fraction1}) and (\ref{fraction2}),
along with $a^\mu_{\lambda,0}=1$ to
calculate the remaining expansion coefficients. Since these
coefficients are the minimal solution, the
coefficients  fall off for large $|r|$. One can then use
(\ref{snormalization}) to find $s^\mu_\lambda(\theta)$. Using
(\ref{sdefinition}) and (\ref{tdefinition}), along with an
algorithm for computing Bessel and Hankel functions, one can
evaluate the spheroidal wave functions. Further details of the
numerical techniques used may be found in Appendix 
\ref{appendixb}.
Figure \ref{FIGURE6}
shows the spheroidal wave functions ${^1 \S^1_2}(z,\theta)$
and ${^2 \S^1_2}(z,\theta)$ as functions of $z$ for
two values of $\theta$.

\section{}
\label{appendixb}

This appendix briefly describes the numerical techniques used to obtain
the results in section \ref{numericalresults}.  One may separate the
problem of numerically evaluating the spheroidal wave functions ${^1
\S^1_2}(z,\theta)$ and ${^2 \S^1_2}(z,\theta)$ into two parts. The
first is to determine the characteristic exponent $\nu^1_2(\theta)$
for a given $\theta$. 
As noted in section \ref{characteristicexponent}
the characteristic exponent $\nu^1_2(\theta)$ is that value of
$\nu$ for which the function $Z^1_2(\theta,\nu)$ vanishes, where
\begin{equation}
Z^1_2(\theta,\nu)\equiv{\cal A}^1_{2,0}(\theta)N^1_{2,-1}(\theta)+
{\cal B}^1_{2,0}(\theta)
+{\cal C}^1_{2,0}(\theta)P^1_{2,1}(\theta).
\label{Zdefinition2}
\end{equation}
To find the root of  $Z^1_2(\theta,\nu)$ (and hence the characteristic
exponent $\nu^1_2(\theta)$ ) one must evaluate the five functions
on the right-hand side of (\ref{Zdefinition2}).
From (\ref{ABCdefinition}) one can see that the three functions 
${\cal A}^1_{2,0},{\cal B}^1_{2,0}$, and ${\cal C}^1_{2,0}$ (here
$\lambda=2$, {\it not} $\nu$) are rational functions of $\nu$
for any $\theta$, and are easily evaluated. The
continued fractions $N^1_{2,-1}(\theta)$ and $P^1_{2,1}(\theta)$,
defined by (\ref{fraction1}) and (\ref{fraction2}), are evaluated
using the modified Lentz's method (see section 5.2 in reference
\cite{PressEtAl}). Both continued fractions usually converge  within 10
iterations when evaluated using this method. The root of 
$Z^1_2(\theta,\nu)$ is found using a simple bisection method.
Although bisection may not be as efficient as other methods, it
has the advantage that it is guaranteed to work once the root
has been bracketed. This is helpful since we are interested in
finding the root $\nu^1_2(\theta)$ of $Z^1_2(\theta,\nu)$ for
many different $\theta$; for some $\theta$ the root $\nu^1_2(\theta)$
lies {\it very} close to a singularity of $Z^1_2(\theta,\nu)$, and
in these cases other root-finding methods may not converge
(see section 9.2 in reference \cite{PressEtAl}). 

Once the characteristic exponent $\nu^1_2(\theta)$ has been calculated,
the remaining problem is to calculate the sums (\ref{newTdefinition}) of expansion coefficients
$a^1_{2,r}(\theta)$ times Bessel functions $J_{\nu+2r+1/2}(2\theta^{1/2}z)$
and $Y_{\nu+2r+1/2}(2\theta^{1/2}z)$. The expansion coefficients are calculated
using the continued fractions $N^1_{2,r}(\theta)$ and $P^1_{2,r}(\theta)$.
With $a^1_{2,0}(\theta)=1$, one has $a^1_{2,1}(\theta)=P^1_{2,1}(\theta)$,
$a^1_{2,2}(\theta)=P^1_{2,2}(\theta)P^1_{2,1}(\theta)$, and in general
\begin{equation}
a^1_{2,n}(\theta)=\prod_{j=1}^n P^1_{2,j}(\theta).
\end{equation}
Likewise, the negative index expansion coefficients are given
by
\begin{equation}
a^1_{2,-n}(\theta)=\prod_{j=1}^n N^1_{2,-j}(\theta).
\end{equation}
The continued fractions $N^1_{2,r}(\theta)$ and $P^1_{2,r}(\theta)$
are calculated using the modified Lentz's method as noted above. The normalization
$s^1_2(\theta)$, defined in (\ref{snormalization}), is calculated at the
same time as the expansion coefficients.

The Bessel functions are calculated most efficiently using recurrence
relations. The functions $J_{\nu+2r+1/2}(2\theta^{1/2}z)$ and $Y_{\nu+2r+1/2}(2\theta^{1/2}z)$ 
with $r\geq 0$ and $r<0$ are handled separately. For $r\geq 0$,  
$J_{\nu+2r+1/2}(2\theta^{1/2}z)$ and $J_{\nu+2r+1/2-2}(2\theta^{1/2}z)$
are calculated for some large $r$ using the routine bessjy() found in
reference \cite{PressEtAl}. Using the recurrence relations for Bessel
functions (see 3.87 and 3.88 in \cite{Jackson}), the $J_{\nu+2r+1/2}(2\theta^{1/2}z)$
are calculated using downward recursion to $r=0$ (this is the direction in which the
recursion is stable for Bessel functions of the first kind \cite{PressEtAl}).
A similar  procedure,  using upward recursion, is used for
the Bessel functions of the second kind $Y_{\nu+2r+1/2}(2\theta^{1/2}z)$. This gives
the necessary Bessel functions for $r\geq 0$. 

The reflection formulae for Bessel functions
are used to calculate the Bessel functions for $r< 0$. Since the index
\begin{equation}
\nu+2r+1/2=-(2|r|-\nu-1/2) \>\>\text{for}\>\>r<0\>\>\text{and}
\>\>1/2<\nu<3/2,
\end{equation}
the same procedure outlined above is used for 
$J_{2|r|-\nu-1/2}(2\theta^{1/2}z)$ and $Y_{2|r|-\nu-1/2}(2\theta^{1/2}z)$,
and then the reflection formulae (see 6.7.19 of \cite{PressEtAl})
\begin{eqnarray}
J_{-\upsilon}(y)&=&\cos\upsilon\pi J_\upsilon(y)-\sin\upsilon\pi Y_\upsilon (y),\\
Y_{-\upsilon}(y)&=&\sin\upsilon\pi J_\upsilon(y)+\cos\upsilon\pi Y_\upsilon (y),
\end{eqnarray}
are used to find 
$J_{\nu+2r+1/2}(2\theta^{1/2}z)$ and $Y_{\nu+2r+1/2}(2\theta^{1/2}z)$ 
for $r< 0$. So all the Bessel functions needed are calculated
with only a few time-consuming calls to the routine bessjy().

Once the expansion coefficients $a^1_{2,r}(\theta)$ 
and the Bessel functions $J_{\nu+2r+1/2}(2\theta^{1/2}z)$
and $Y_{\nu+2r+1/2}(2\theta^{1/2}z)$ are tabulated, the sums 
in (\ref{newTdefinition}) are calculated. Care must be taken when
terminating the sums. Although the expansion coefficients
$a^1_{2,r}(\theta)$ fall off {\it very} quickly as $|r|$ becomes
large, the products $a^1_{2,r}(\theta)J_{\nu+2r+1/2}(2\theta^{1/2}z)$
and $a^1_{2,r}(\theta)Y_{\nu+2r+1/2}(2\theta^{1/2}z)$ may not
fall off as fast. We terminate the sums when
the {\it products} of the expansion coefficients and Bessel functions
no longer contribute (at double-precision accuracy) to the sums.

The primary numerical technique used to evaluate the multipole moments
(\ref{moments}) is numerical integration. Both the integral over
$\kappa$ in (\ref{moments}) and the integral over $x$ in
(\ref{Gldefinition}) were done using a fifth order embedded
Runge-Kutta-Fehlberg algorithm with adaptive stepsize control
\cite{PressEtAl}.  Although formally the upper limit of the integral
over $\kappa$ extends to infinity, we only integrated until the
remaining contribution became negligible. The Bessel function with
index $l+1/2$ was evaluated with the routine bessjy().



\begin{figure}
\caption{Multipole moments $\langle a_l^2\rangle$ normalized to
the quadrupole $\langle a_2^2\rangle$. The horizontal axis is
the index $l$ of the multipole moment and the vertical axis
is $M_l$. See (\protect\ref{Mldefinition}) for the definition
of $M_l$. The points labeled ``mixed''
show results obtained from the (\protect\ref{moments}) 
and (\protect\ref{Gldefinition}) which analytically model a universe
containing both dust and radiation, and which is not completely
dust dominated at last-scattering.
The points labeled ``transfer function''
show results from Turner, White, and Lidsey \protect\cite{Turner2} obtained
using a transfer function.}
\label{FIGURE1}
\end{figure}

\begin{figure}
\caption{Multipole moments $\langle a_l^2\rangle$ normalized to
the quadrupole $\langle a_2^2\rangle$. The axes are the same
as in Figure \protect{\ref{FIGURE1}}.
The points labeled ``mixed''
show results obtained from (\protect\ref{moments}) 
and (\protect\ref{Gldefinition}) which analytically model a universe
containing both dust and radiation, and which is not completely
dust dominated at last-scattering.
The points labeled ``exact-dust''  show
results obtained using Eq. (6.2) of Allen and Koranda \protect\cite{AllenKoranda}.
The points labeled ``approximate-dust'' shows results obtained by evaluating Eq. (6.1) (which is
a long-wavelength approximation to Eq. (6.2) ) of Allen
and Koranda \protect\cite{AllenKoranda}. Both 
sets of points ``exact-dust'' and ``approximate-dust''
are for universe models that are {\it completely} dust dominated at
last-scattering.}
\label{FIGURE2}
\end{figure}

\begin{figure}
\caption{Multipole moments $\langle a_l^2\rangle$ normalized to
the quadrupole $\langle a_2^2\rangle$.  The axes are the same
as in Figure \protect{\ref{FIGURE1}}.  
The points labeled ``mixed''
show results obtained from (\protect\ref{moments}) 
and (\protect\ref{Gldefinition}) which analytically model a universe
containing both dust and radiation, and which is not completely
dust dominated at last-scattering.
The points labeled ``Boltzmann'' show  results from Dodelson, Knox, and Kolb
\protect\cite{KnoxDodelson}, who do not use the Sachs-Wolfe
formula to calculate the CBR anisotropy.  Instead they use numerical
methods to evolve the photon distribution function using first-order 
perturbation theory of the general relativistic Boltzmann equation for
radiative transfer.}
\label{FIGURE3}
\end{figure}

\begin{figure}
\caption{The function $Z^1_2(\theta,\nu)$ plotted versus $\nu$
for  $\theta=5$.
The value of $\nu$ at which this function vanishes
is the characteristic exponent $\nu^1_2(\theta)$
for  $\theta=5$. This plot is typical for 
values of $\theta$ in the range $10^{-4}\leq\theta\leq 10^4$.
The root is always located between two singularities.}
\label{FIGURE4}
\end{figure}

\begin{figure}
\caption{The characteristic exponent $\nu^1_2(\theta)$
plotted vs $\log\theta$. Only for  certain discrete values of
$\theta$ is $\nu$ equal to $1$. For this reason
the function \protect{ $S^{1(j)}_1(z,\theta)$} is not a valid
solution to the differential equation (\protect{\ref{diffeq}})
for $\mu=1,\lambda=2$, and $\theta$ arbitrary.}
\label{FIGURE5}
\end{figure}

\begin{figure}
\caption{The spheroidal wave functions
${^1 \S^1_2}(z,\theta)$ (solid lines) and ${^2 \S^1_2}(z,\theta)$
(dotted lines) for $\theta=0.1$ and $\theta=10$. Both functions
diverge in the limit as \protect{$z\rightarrow 1$}. Note that the vertical
scales are different for the two plots. }
\label{FIGURE6}
\end{figure}

\begin{table}
\caption{List of free parameters that define the cosmological model}
\begin{tabular}{cccl}
Parameter & Units &Range & Description\\
\tableline
&&&\\ 
$H_0$ &  $\rm{length}^{-1}$&$ H_0>0$ & Present-day Hubble expansion rate\\
$ Z_{\rm ls}$ & dimensionless &$\zls>0$& Redshift at last scattering of CBR\\
$ Z_{\rm eq}$ & dimensionless& unrestricted & Redshift at equal matter/radiation energy density \\
$ Z_{\rm end}$ & dimensionless&$\zend>\zeq,\zls$ & Redshift at end of de Sitter inflation\\
\end{tabular} 
\label{table1}
\end{table}

\begin{table}
\caption{Multipole moments $\langle a_l^2\rangle$ evaluated using
different methods. These have been divided by the scale of the moments
$\rho_{\rm ds}/ \rho_{\rm p}$.
The second column shows multipoles for a mixed cosmology 
obtained from the formulae (\protect\ref{moments}) and
(\protect\ref{Gldefinition}), which analytically model a universe
containing both dust and radiation, and which is not completely
dust dominated at last-scattering. The third column shows results
obtained by evaluating Eq. (6.2) of Allen and Koranda
\protect{\cite{AllenKoranda}}, which assumes that the universe
was completely dust dominated at last-scattering. The fourth column shows
results obtained by evaluating Eq. (6.1) of Allen and Koranda,
which is a long wavelength approximation to their Eq. (6.2),
and equivalent to standard formula of Abbott and Wise \protect\cite{AbbotWise}
and Starobinsky \protect\cite{Starobinsky}. The fifth column shows results
obtained by Ng and Speliotopoulos \protect\cite{Ng}, who numerically integrated
the wave equation to obtain the amplitude of the gravitational waves. They
also considered a universe model which is not completely dust dominated
at last-scattering. All 
the results were obtained with $Z_\protect{\rm eq}=6000$ 
and $Z_\protect{\rm ls}=1100$.}
\begin{tabular}{ccccc}
$l$&$\langle a_l^2\rangle{\rho_{\rm P}\over \rho_{\rm dS}}$
&$\langle a_l^2\rangle{\rho_{\rm P}\over \rho_{\rm dS}}$
&$\langle a_l^2\rangle{\rho_{\rm P}\over \rho_{\rm dS}}$&
$\langle a_l^2\rangle{\rho_{\rm P}\over \rho_{\rm dS}}$\\
&this work 
& Allen, Koranda & Allen, Koranda 
& Ng, Speliotopoulos\\
&Eq. (\ref{moments})&Eq. (6.2) of ref. \protect\cite{AllenKoranda}&Eq. (6.1)  of ref. 
\protect\cite{AllenKoranda}
&Table 1 of ref. \protect\cite{Ng}\\
\tableline
&&&&\\
2&1.54&1.52&1.52&1.55\\
3&6.07$\times 10^{-1}$&6.07$\times 10^{-1}$&6.07$\times 10^{-1}$&\\
4&3.44$\times 10^{-1}$&3.44$\times 10^{-1}$&3.44$\times 10^{-1}$&\\
5&2.27$\times 10^{-1}$&2.27$\times 10^{-1}$&2.27$\times 10^{-1}$&\\
6&1.63$\times 10^{-1}$&1.62$\times 10^{-1}$&1.62$\times 10^{-1}$&\\
7&1.23$\times 10^{-1}$&1.22$\times 10^{-1}$&1.22$\times 10^{-1}$&\\
8&9.66$\times 10^{-2}$&9.61$\times 10^{-2}$&9.58$\times 10^{-2}$&\\
9&7.80$\times 10^{-2}$&7.74$\times 10^{-2}$&7.71$\times 10^{-2}$&\\
10&6.43$\times 10^{-2}$&6.37$\times 10^{-2}$&6.34$\times 10^{-2}$&\\
20&1.74$\times 10^{-2}$&1.69$\times 10^{-2}$&1.66$\times 10^{-2}$&1.75$\times 10^{-2}$\\
50&2.36$\times 10^{-3}$&1.99$\times 10^{-3}$&1.83$\times 10^{-3}$&2.36$\times 10^{-3}$\\
100&1.91$\times 10^{-4}$&8.67$\times 10^{-5}$&6.23$\times 10^{-5}$&1.90$\times 10^{-4}$\\
150&7.45$\times 10^{-6}$&4.11$\times 10^{-6}$&1.89$\times 10^{-6}$&7.41$\times 10^{-6}$\\
200&4.56$\times 10^{-6}$&3.68$\times 10^{-6}$&9.96$\times 10^{-7}$&4.49$\times 10^{-6}$\\
\end{tabular}
\label{table2}
\end{table}


\begin{references}
\bibitem{AllenKoranda}
Bruce Allen and Scott Koranda, Phys. Rev. D {\bf 50}, 3713 (1994).
Note that the caption for Table I of this reference should indicate
that the multipoles listed in the table
have been divided by $\rho_{\rm ds}/\rho_{\rm P}$. Also note that the
definition of $M_l$ given in the caption of Figure 2 contains
a spurious factor of $\rho_{\rm ds}/ \rho_{\rm p}$;
the correct definition is given by (\protect\ref{Mldefinition})
of this paper.
\end{references}
\end{document}